\documentclass[12pt]{article}
\usepackage{epsf}
\usepackage{rotating}

\let\section=\subsection     \let\subsection=\subsubsection                                           \usepackage{epsfig}

\begin{document}

\begin{center}
{\large \textbf{SELECTED RESULTS FROM}}\\[2mm]
{\large \textbf{GROUND-BASED COSMIC RAY AND}}\\[2mm]
{\large \textbf{GAMMA-RAY EXPERIMENTS\footnote{Invited talk at
'Nuclear Astrophysics', the International Workshop XXVI 
on Gross Properties of Nuclei and 
Nuclear Excitations, Hirschegg, Austria, January 1998.}}}\\[5mm]
N.~MAGNUSSEN \\[5mm]
{\small \textit{Bergische Universit\"at Wuppertal \\[0pt]
Gau\ss str. 20, D-42097 Wuppertal, Germany}}
\end{center}

\begin{abstract}
\noindent Selected results from the HEGRA experiment on charged Cosmic Rays
and on very high energy gamma-rays are presented.
The MAGIC Teles\-cope is presented as an outlook to the future of Gamma-Ray
astronomy.
\end{abstract}

\section{Introduction to Ground-Based Cosmic Ray Data}

As a general rule the dynamic range of precision detectors is limited to
roughly 2 to 3 orders of magnitude in energy. In case of the charged Cosmic
Rays (CR) with an energy spectrum extending over more than 13 orders of
magnitude, this neccesitates a large number of different experimental setups
in order to cover the full spectrum. Space-borne, i.e. direct, experiments,
cover the spectrum from $\approx $ 10$^{7}$ eV to $\approx $ 10$^{15}$
eV/nucleon, and ground-based experiments operate above total energies of a
few 10$^{12}$ eV up to more than 10$^{20}$ eV.

In the following we will concentrate on the ground-based measurements. Here
various experiments which are sensitive in the energy region around 10$^{15}$
eV consistently show a significant steepening of the all-particle spectrum
around this energy. When studying the data more closely, however, the
agreement between the experiments turns out to be not so good, i.e.,
differences well above the fluctuations given by the individual errors. This
is shown in fig.~1 where the data on the 'knee' in the all-particle spectrum
are collected \cite{bws}. From these data one must conclude that the
absolute position, the 'sharpness' of the knee, and also the absolute flux
in this energy region are more uncertain than expected from the individual
errors. 
\leavevmode
\begin{figure}[tbp]
\centering \leavevmode
\epsfxsize=10cm \epsffile{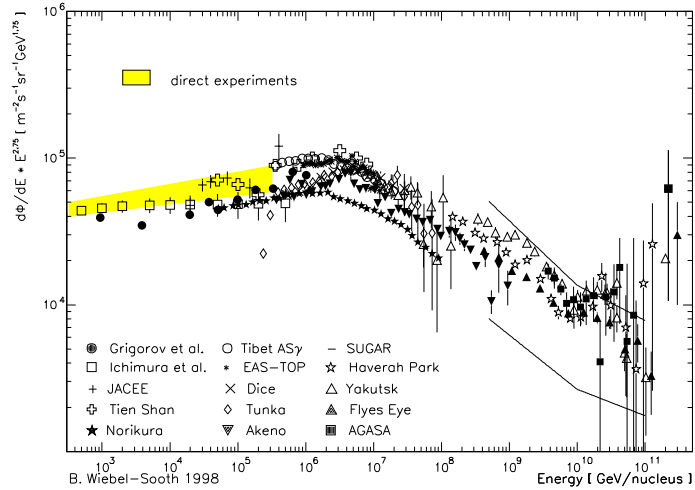}
\caption{{\protect\small The charged Cosmic Rays all-particle spectrum
around the 'knee' as measured by a number of ground-based experiments (taken
from \protect\cite{bws}).}}
\label{fig-1}
\end{figure}
Ground-based experiments use a detector, i.e. the atmosphere as absorber
with some added readout elements, like scintillators, Cherenkov detectors,
etc., which can only be calibrated in the laboratory to a very limited
degree. The calibration therefore has to rely very heavily on MC simulations
of the development of the extensive air showers and of the performance of
the detectors. One possible reason for the deviations in the measured
spectra might thus be the use of different Monte Carlo (MC) generators in
the data analyses. This will be briefly discussed below for the most recent
data.

\section{The Energy Spectrum and the Chemical Composition}

Currently the two experiments HEGRA \cite{hegra}, located within the
Observatorio del Roque de los Muchachos on the Canary island La Palma, and
KASCADE \cite{kaskade}, located near Karlsruhe, Germany, are analysing their
airshower data with respect to the all-particle spectrum and the chemical
composition of the CR near the 'knee'. In the following we shall discuss
some of the preliminary HEGRA results and compare to KASCADE where
appropriate.

The HEGRA setup for the detection of CR with energies above $\approx$ 15 TeV
has 3 components: (i) 243 scintillator stations of 1 m$^2$ each distributed
over 40,000 m$^2$, (ii) 77 AIROBICC Cherenkov detectors spread out over the
same area, and (iii) 17 Geiger towers within the central 15,000 m$^2$. 

\subsection{The Reconstruction of Energy and Mass}

The measured distribution of particle densities in the scintillator array is
fitted to yield the shower size N$_{e}$at detector level and the correlated
shower size at the maximum (X$_{\max }$) of the shower development. 
The lateral Cherenkov light density, $\rho _{C}(r)$, is analysed in the
intervall 7.5 $<r<$ 100 m and can be described by an exponential 
\begin{equation}
\rho _{C}(r)=a\cdot exp(r\cdot R_{light}),
\end{equation}
where the shape parameter, $R_{light}$, is inversely proportional to the 
\textit{absolute} penetration depth, X$_{max}$, of the analysed shower. As
MC investigation show, this proportionality is essentially independent of
the mass, A, of the incident primary particle. The achieved resolution for X$%
_{max}$ varies between 1.0 and $\approx $ 0.5 radiation lenghts (X$_{0}$ =
36 g/cm$^{2}$) for energies between 100 TeV and a few PeV. The
reconstruction of the energy in two independent analyses relies either on a
combination of N$_{e}$ and $R_{light}$, or on a combination of N$_{e}$, N$%
_{\mu }$(the corresponding muon shower size), and $R_{light}$. In both cases
an energy resolution of $\approx $30\% is achieved. Using the CORSIKA
simulation program \cite{corsika} different hadronic interaction generators
(i.e., VENUS, QGSJET, HDPM, DPMJET, SIBYLL) can be used. MC investigation
have shown that the qualitative sensitivity for physics parameters and e.g.
the energy resolution is independent of the hadronic interaction generator.
The absolute energy scale, however, turns out to be dependent on the MC
generator and the chemical composition.

\subsection{The Energy Spectrum}

Using the VENUS model within CORSIKA, HEGRA performed a regularized
unfolding of the measured N$_e$ distributions on a run-by-run basis (see 
\cite{bws} for details). The result of the unfolding is shown in fig.~2. 
\leavevmode
\begin{figure}[t]
\centering \leavevmode
\epsfxsize=10cm \epsffile{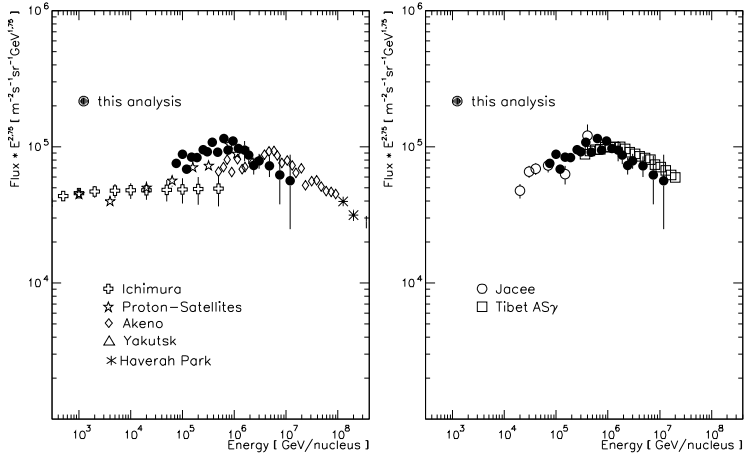}
\caption{{\protect\small The charged Cosmic Rays all-particle spectrum
around the 'knee' as measured by HEGRA compared to older and recent (Tibet)
data of other experiments. The unfolding of the detector mearing was done by
a regularized unfolding procedure based on Monte Carlo events simulated with
the VENUS interaction generator. (for details see \protect\cite{bws}).}}
\label{fig-2}
\end{figure}

The spectral index of the all-particle spectrum \textit{below} the 'knee' is
found to be 
\begin{equation}
\gamma = 2.63 \pm 0.02 \mathrm{(stat.)} \pm 0.05 \mathrm{(syst.)} \pm 0.05 
\mathrm{(model)},
\end{equation}
where the last error is an estimation of the dependence on the hadronic
interaction generator in the MC.

\subsection{The Chemical Composition around the 'Knee'}

In order to get a first indication of the behaviour of the chemical
composition around the 'knee' and in order to check the MC generators, HEGRA
made in one analysis the simple ansatz of using parametrized MC
distributions for the energy determination \cite{lindner}. A more involved
unfolding procedure based on the full MC and detector information is used in
a second analysis decribed in detail in \cite{bws}. In both analyses the
composition around the 'knee' is observed to be consistent with an unchanged
or slightly heavier composition. For illustrating the results and the
observed MC dependence the first approach is discussed in the following.

In the energy range from 300 TeV to 3 PeV the height of the individual
shower maxima can be reconstructed with an accuracy comparable to the RMS
spread for iron showers (rms(X$_{max},_{iron}$) = 33 $\pm $ 2 g/cm$^{2}$)
and thus much better than the RMS spread for proton showers (rms(X$%
_{max},_{proton}$) = 84 $\pm $ 6 g/cm$^{2}$). Since the mean X$_{max}$
values of protons and iron showers differ by about 140 g/cm$^{2}$, the
fractions of light (hydrogen and helium) and heavy elements (oxygen and
iron) contributing to the measured distribution can thus be obtained from
fitting model X$_{max}$distributions to the data for different energy bins.
The details of this analysis can be found in \cite{cortina}. Under the
assumption that the chemical composition is the same for the different
energy bins, the mean fraction of light elements is fitted to be 0.52 $\pm $
0.10 (total error), which within errors, is compatible with direct
measurements below 100 TeV. 

Besides the shape of the X$_{max}$ distribution the mean value may serve to
determine the elemental composition. The measured mean values, however, turn
out to be systematically smaller by 45 $\pm $ 24 g/cm$^{2}$ than expected
from MC for the composition compatible with the fitted fractions which are
based on the \textit{detailed shape} of the X$_{max}$ distribution. This
apparant discrepancy between data and MC can be due to detector,
atmospheric, or MC generator effects. After careful investigations HEGRA now
suspects the longitudinal shower development in the MC to yield too large X$%
_{max}$ values due to a non-perfect simulation of the longitudinal shower
development. The investigations, however, are not yet completed. Note that
the AIROBICC-type detectors as employed e.g. by HEGRA are the only airshower
detectors which are sensitive to the \textit{absolute} position of the
shower maximum due to the translation of the longitudinal development into
the radial Cherenkov photon density. This is a consequence of the varying
refractive index and height of emission along the shower.

Like HEGRA, the KASCADE collaboration also finds little change in the
chemical composition around the 'knee' \cite{knapp}. The absolute
determination which relies on the MC is, however, still biased by large
discrepancies between some of the hadronic interaction generators. Here
still considerable effort is needed in order to achive reliable results in
the future.

\section{Gamma-Ray Astronomy}

Since the detection of the Crab nebula as a source of gamma-rays ($\gamma $%
-rays) with energies above 500 GeV in 1989 \cite{whipple}, about 10 galactic
and extragalactic sources of very high energetic photons (E $>$ 300 GeV)
have been detected by about as many experiments. The status of the field is
indicated in fig. 3 which shows the skymap of the very high energy (VHE),
i.e., $E >$ 300 GeV, $\gamma $-ray sources as of August 1997.
Note that contrary to former beliefs also extragalactic sources of VHE $%
\gamma $-rays were discovered during the last 5 years. The fast progress of
the field will be illustrated by the most recent HEGRA results concerning
the brightest source on the $\gamma $-ray sky in 1997, the active galactic
nucleus (AGN) Mkn 501 at a distance of $z$ = 0.034 ($\approx $ 600 million
light years, for H$_{0}$= 50 km sec$^{-1}$Mpc$^{-1}$).
\leavevmode
\begin{figure}[t]
\centering \leavevmode
\epsfxsize=10cm \epsffile{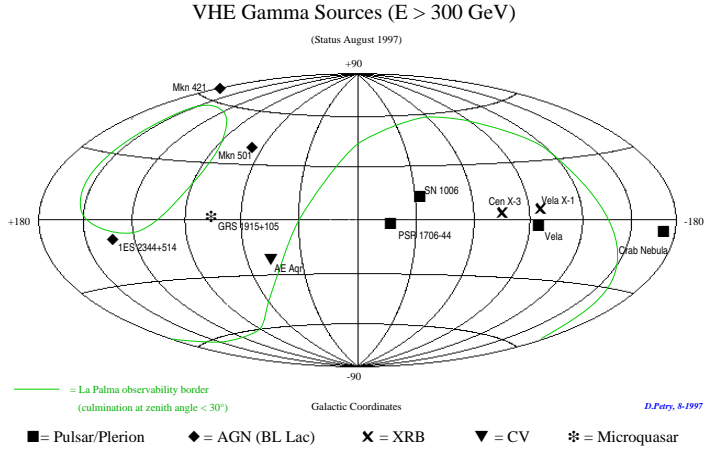}
\caption{{\protect\small The galactic coordinates of point sources of
Gamma-radiation of energy $E >$ 300 GeV as observed by the various Cherenkov
telescope observatories around the world \protect\cite{petry}.}}
\label{fig-3}
\end{figure}

\subsection{The HEGRA Cherenkov Telescopes}

In 1996 the HEGRA experiment completed the installation of six Imaging Air
Cherenkov Telescopes (IACTs). Four identical IACTs are operated in
coincident mode, i.e. for each airshower four different views are recorded.
Hereby the energy threshold can be lowered and the $\gamma $/hadron
separation efficiency can be raised. This is an advantage for the study of
weak and especially for extended sources but is less than optimal for strong
point-like sources due to the restricted effective collection area. For the
current energy range of IACTs, i.e. above 300 GeV, and the current stage of $%
\gamma $-ray astronomy, i.e. the discovery age in which we learn how to
optimize the detectors, this setup, however, has its merits due to the
redundancy of the information. For energies below 100 GeV where air showers
only produce very little Cherenkov light, however, the redundancy will be
lost to a large degree due to the very limited number of particles above the
Cherenkov Threshold, i.e., different telescopes will view different tracks
in the shower. In this energy regime therefore very sensitive telescopes
like the MAGIC Telescope discussed below will be needed and which will not
compromise on the effective photo collection area.

\subsection{The Extragalactic TeV photon source Mkn 501}

Mkn 501 belongs to the blazar class of AGN, i.e. it is an AGN with a large
plasma jet which is pointing along the line-of-sight, and it has been
observed at TeV energies since 1995 \cite{whipple2} when its activity level
corresponded to about 8\% of the flux of the strongest galactic source, the
Crab nebula. Due to its steady emission level the Crab nebula has become the
Standard Candle for $\gamma $-ray astronomy. The activity level of Mkn 501
in 1996 rose to about 30\% of the Crab flux \cite{bradbury} and in 1997
HEGRA recorded on average a flux corresponding to about 200\% of the Crab
flux, and, in addition to this much higher average activity level, huge
flares on time-scales as short as a day or less were recorded. As summarized
in \cite{protheroe} 
a great number of Cherenkov telescopes (Whipple, CAT, TACTIC, Telescope
Array, HEGRA) 
were able to detect this signal and record a light curve. \leavevmode
\begin{figure}[tbp]
\centering \leavevmode
\begin{turn}{270}
\epsfxsize=7.5cm
\epsffile{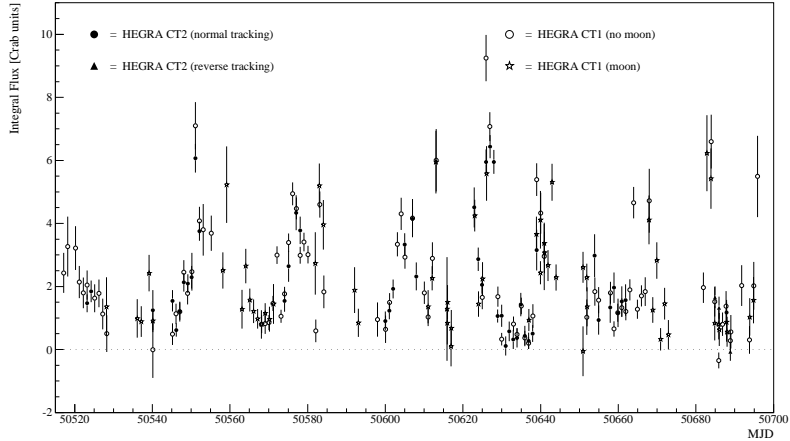}
\end{turn}
\caption{The light curve of the blazar Mkn 501 as measured by the HEGRA
telescopes CT1 and CT2 during the 1997 observation campaign. MJD (Modified
Julian Date) 50700 corresponds to the 8th of September 1997.}
\label{fig-4}
\end{figure}
%
\leavevmode
\begin{figure}[tbp]
\centering \leavevmode
\epsfxsize=6cm \epsffile{mkn501_1997.eps}
\caption{The energy spectrum of the blazar Mkn 501 as measured by HEGRA
system telecopes during the 1997 observation campaign. Measured points beyond
10 TeV are not shown.}
\label{fig-5}
\vspace{0.5cm}
\centering \leavevmode
\epsfxsize=8.5cm \epsffile{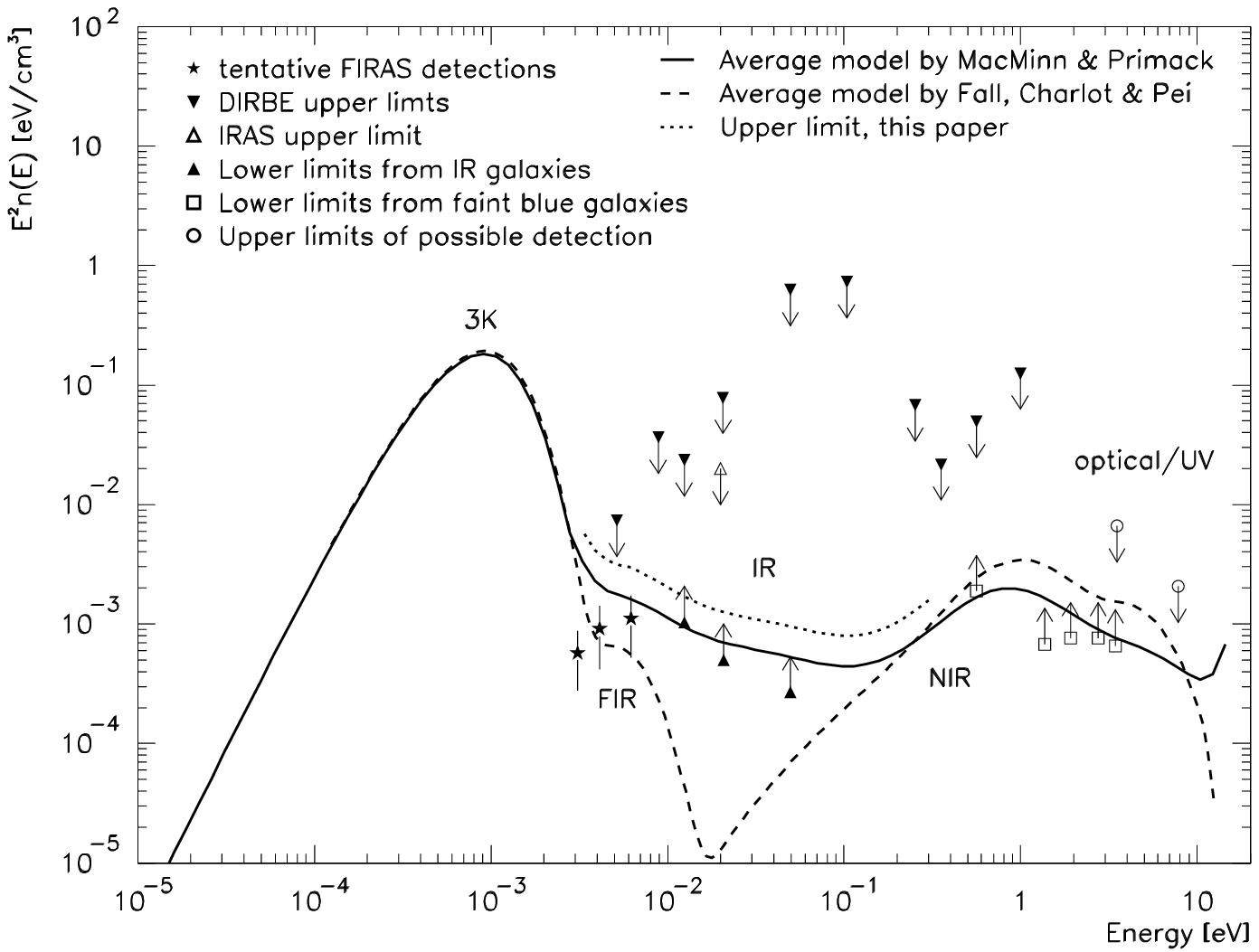}
\caption{Energy density of the extragalactic diffuse background radiation.
For detailed references see \protect\cite{funk}. 
\textit{Solid line:} average model from MacMinn
\& Primack including the CMBR \protect\cite{MacMinn}, \textit{dashed line:}
average model from Fall, Charlot \& Pei added to the CMBR \protect\cite{fall}%
, \textit{dotted line:} upper limit derived in \protect\cite{funk}.}
\label{fig-6}
\end{figure}
The most complete light curve was obtained by the stand-alone telescopes CT1
and CT2 of the HEGRA collaboration, because CT1 was also operational during
moonshine. The light curve extending from March until September 1997 is
shown in fig.~4. The short-term flares point towards very compact sources,
i.e. regions in the vicinity of the supermassive ($\mathcal{O}$(10$^{8}$)M$%
_{\odot }$) black hole thought to be in the centre of the galaxy and
powering the very strong non-thermal emission of the AGN. These large
fluxes, which for the first time were recorded from an extragalactic source,
allowed the determination of the energy spectrum extending beyond 5 TeV. The
energy spectrum as measured by the HEGRA IACT system telescopes in 1997 is
shown in fig.~5 \cite{aharonian}. As the analysis of systematic effects at
higher energies is not yet completed, the spectrum is only shown up to 10
TeV.

\begin{figure}[thb]
\centering \leavevmode
\epsfxsize=7cm
\hspace{-8cm}
\begin{rotate}{-90}
\epsffile{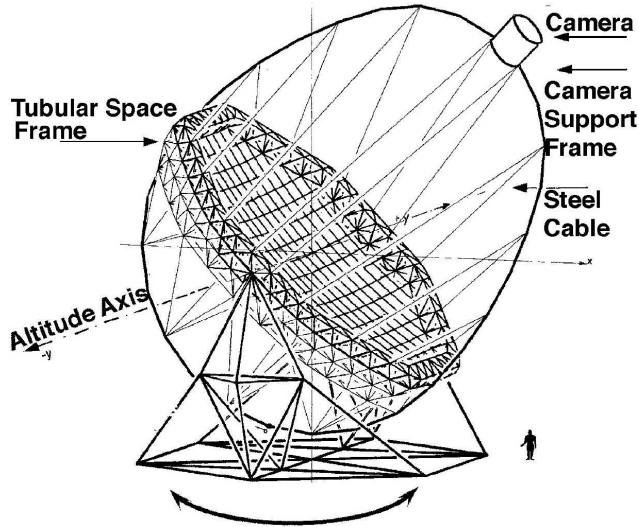}
\end{rotate}
\vspace{6.5cm}
\caption{Sketch of the planned 17 m diameter MAGIC Telescope \protect\cite
{magic}.}
\label{fig-7}
\end{figure}

But already the unabsorbed spectrum extending up to 10 TeV allows to extract
a fundamentally important upper limit on the density of infrared (IR) relic
photons in intergalactic space. This density is dependent on the era of
galaxy formation and thus on the nature of the Dark Matter.

One extraction of an upper limit on the infrared photon density was
performed in \cite{funk} where it is assumed that the IR background is
uniformly distributed in the Universe. The result, together with existing
upper limits, lower limits, and an average model for galaxy evolution based
on a mixed Dark matter ansatz is shown in fig.~6. The resulting low density
of infrared photons limits the parameter space for Dark Matter models and at
the same time opens up the deeper regions of the universe (i.e. $z$ = $%
\mathcal{O}$(0.1)) for TeV astronomy.

\subsection{The Future of Gamma-Ray Astronomy}

Currently space-borne $\gamma $-astronomy is limited to energies \textit{%
below} 10 GeV and ground-based $\gamma $-astronomy to energies \textit{above}
300 GeV. In both cases the reasons are limitations of sensitivity and/or
effective collection areas. The result is an observation gap between 10 GeV
and 300 GeV where the Universe has not been observed in and where we expect
to find hints or answers to important physics questions in astrophysics,
cosmology, and particle physics.

In order to bridge this gap the ground-based 17 m diameter MAGIC Telescope
(see fig.~7) has been designed during the last 2 years \cite{magic}. Using
innovative elements it will be possible to close the last observation gap
for about 1\% (!) of the cost of a satellite experiment, which until now was
believed to be necessary in order to do measurements in this energy domain.
At the same time the sensitivity in the energy region of current Cherenkov
telescopes will be improved by up to an order of magnitude.

\end{document}